\documentclass[twocolumn,showpacs,preprintnumbers,amsmath,amssymb]{revtex4}

\usepackage{graphicx}

\begin{document}

\title{
Deflection angle of light in an Ellis wormhole geometry
}
\author{Koki Nakajima}
%\email{}
\author{Hideki Asada} 
%\email{asada@phys.hirosaki-u.ac.jp}
\affiliation{
Faculty of Science and Technology, Hirosaki University,
Hirosaki 036-8561, Japan} 

\date{\today}

\begin{abstract}
We reexamine the light deflection by an Ellis wormhole. 
The bending angle as a function of the ratio between 
the impact parameter and the throat radius of the wormhole 
is obtained in terms of a complete elliptic integral 
of the first kind. 
This result immediately yields asymptotic expressions 
in the weak field approximation. 
It is shown that an expression for the deflection angle 
derived (and used) in recent papers is valid at the leading order 
but it breaks down at the next order because of 
the nontrivial spacetime topology. 
\end{abstract}


\maketitle

\section{Introduction}
%\noindent \emph{Introduction.--- } 
The bending of light was the first experimental confirmation of 
the theory of general relativity. 
At present, the gravitational lensing 
is one of the important tools in astronomy and cosmology. 
It is widely used for investigating extrasolar planets, 
dark matter and dark energy. 

The light bending is also of theoretical importance, 
in particular for studying a null structure of a spacetime. 
For example, 
strong gravitational lensing in a Schwarzschild black hole 
was considered by Frittelli, Kling and Newman \cite{Frittelli} 
and by Virbhadra and Ellis \cite{VE2000};  
Virbhadra and Ellis \cite{VE2002} later described 
the strong gravitational lensing by naked singularities; 
Eiroa, Romero and Torres \cite{ERT} treated 
Reissner-Nordstr\"om black hole lensing. 
 
A peculiar feature of general relativity is that 
the theory admits a nontrivial topology of a spacetime, 
for instance a wormhole. 
An Ellis wormhole is a particular example of the Morris-Thorne 
traversable wormhole class \cite{Ellis, Morris1, Morris2}.
Many yeas ago, scattering problems in such spacetimes were discussed 
(for instance, \cite{CC, Clement}). 
One remarkable feature is that the Ellis wormhole has a zero mass 
at the spatial infinity but it causes the light deflection 
\cite{CC, Clement}. 
Moreover, the gravitational lensing by wormholes has been recently 
investigated as an observational probe of such an exotic spacetime 
\cite{Safonova, Shatskii, Perlick, Nandi, Abe, Toki}. 
Perlick \cite{Perlick}, Nandi, Zhang and Zakharov \cite{Nandi}, 
Dey and Sen \cite{DS} calculated 
a deflection angle of light due to an Ellis wormhole, 
though their expressions are in different forms. 
Therefore, a reason for such differences should be clarified. 

Moreover, a rigorous form of the bending angle plays an important role 
in understanding properly a strong gravitational field  
\cite{Frittelli, VE2000, VE2002, Perlick}. 
The main purpose of this brief paper is 
to reexamine the bending angle of light by the Ellis wormhole 
in order to clarify an unclear relationship 
among the different expressions. 
We shall show that 
the deflection angle as a function of the impact parameter 
and the throat radius of the wormhole is obtained 
in terms of a complete elliptic integral of the first kind. 
We discuss also the validity and limitation of 
several forms of the deflection angle by wormholes, 
which have been recently derived and often used 
\cite{Perlick, Nandi, DS, BP, Abe, Toki}. 
We take the units of $G=c=1$ throughout this paper.

\section{Deflection angle of light by the Ellis wormhole}
The line element for the Ellis wormhole is written as 
\cite{Ellis, Perlick, Nandi}
\begin{equation}
ds^2 = - dt^2 + dr^2 + (r^2 + a^2) (d\theta^2 + \sin^2\theta d\phi^2) . 
\label{ds}
\end{equation}
To cover the entire wormhole geometry, the coordinate $r$ 
runs from $-\infty$ to $+\infty$, 
where $r=0$ corresponds to the throat of the wormhole. 
In order to discuss the deflection angle of light, 
it is sufficient to consider 
$r \in (0, +\infty)$, only one half of the wormhole geometry. 
This metric gives the Lagrangian for a massless (light-like) particle as 
\begin{equation}
L = - \dot{t}^2 + \dot{r}^2 
+ (r^2 + a^2) (\dot{\theta}^2 + \sin^2\theta \dot{\phi}^2) , 
\label{L}
\end{equation}
where the dot denotes the derivative with respect to 
the affine parameter. 

The Ellis wormhole is spherically symmetric 
so that a photon orbit can be considered 
on the equatorial plane $\theta = \pi/2$ 
without loss of generality. 
Since this spacetime is stationary and spherically symmetric, 
we have two constants of motion of a photon as 
\begin{eqnarray}
E &\equiv& \dot{t} , 
\label{E}
\\
h &\equiv& (r^2+a^2)\dot{\phi} , 
\label{h}
\end{eqnarray} 
where $E$ and $h$ are corresponding to the photon's specific energy 
and the photon's specific angular momentum, respectively. 
The two constants of motion are substituted into the null condition 
$ds^2 = 0$ 
to obtain an equation for the photon orbit as 
\begin{equation}
\frac{1}{(r^2+a^2)^2} \left( \frac{dr}{d\phi} \right)^2 
= \frac{1}{b^2} - \frac{1}{r^2+a^2} , 
\label{orbit}
\end{equation}
where a constant $b$ is defined as $h/E$. 
The impact parameter is the perpendicular coordinate distance 
between the projectile's fiducial path and the center of a deflector 
by assuming that the fiducial path were not deflected. 
For the Ellis wormhole case, the zero deflection limit 
is obtained by $a \to 0$. 
If $a = 0$, $r=b$ means that $r$ is the minimum 
according to Eq. (\ref{orbit}). 
Namely, the above constant $b$ can be called the impact parameter 
of the light trajectory. 
On the other hand, the closest approach $r_0$ 
between the light trajectory and the coordinate origin (the deflector) 
is given by Eq. (\ref{orbit}) as 
\begin{equation}
r_0 = \sqrt{b^2 - a^2} .
\label{r0}
\end{equation} 
Namely, $r_0$ is the minimum value of the radial coordinate 
along the light ray. 

An integration of Eq. (\ref{orbit}) immediately gives 
the deflection angle expressed as 
\begin{equation}
\alpha(b) = 2 \int_{r_0}^{\infty} 
\frac{bdr}{\sqrt{(r^2+a^2)^2 - (r^2+a^2)b^2}} 
- \pi . 
\label{alpha-r}
\end{equation}
We make a coordinate transformation from $r \in [0, +\infty)$ 
to $R \in [a, +\infty)$ 
by $R^2 = r^2+a^2$, where 
$R$ is the circumference radius. 
Eq. (\ref{alpha-r}) becomes 
\begin{equation}
\alpha(b) = 2 \int_{b}^{\infty} 
\frac{bdR}{\sqrt{(R^2-a^2) (R^2-b^2)}} 
- \pi . 
\label{alpha-R}
\end{equation}
This is rewritten as 
\begin{eqnarray}
\alpha(b) &=& 2 \int_0^1 \frac{dt}{\sqrt{(1-t^2)(1 - k^2 t^2)}} 
- \pi 
\nonumber\\
&=& 2 K(k) - \pi , 
\label{alpha2}
\end{eqnarray}
where $t \equiv b/R$ and $k \equiv a/b$. 
The integral in Eq. (\ref{alpha2}) 
is a complete elliptic integral of the first kind $K(k)$, 
which admits a series expansion for $k <1$. 
Hence, Eq. (\ref{alpha2}) is expanded as 
\begin{equation}
\alpha(b) = \pi \sum_{n=1}^{\infty} 
\left[ \frac{(2n-1)!!}{(2n)!!} \right]^2 k^{2n} . 
\label{alpha3}
\end{equation}

\section{Comparison with previous results} 
Perlick \cite{Perlick} and Nandi, Zhang and Zakharov \cite{Nandi}
later obtained the deflection angle in a different form 
(e.g., Eq. (54) in \cite{Nandi}) 
that is expressed in terms of the closest approach 
\cite{Perlick, Nandi}. 
It follows that their expression using the closest approach 
can be recovered from Eq. (\ref{alpha2}) 
by noting $r_0^2 = b^2 - a^2$ \cite{Nandi-2}. 
However, the present result by Eq. (\ref{alpha2}) 
is more convenient for astronomers, 
especially on a microlens study, 
since describing an image direction (its angular position) 
needs the impact parameter rather than the closest approach. 

Dey and Sen \cite{DS} followed the method proposed 
by Amore and Arceo \cite{Amore, Amore2}, 
in which firstly the linear delta function technique is used 
to approximate the above type of the integral with 
an {\it ansatz} potential and next 
the principle of minimal sensitivity (PMS) is used to 
minimize the parametric dependence on the deflection angle. 
They obtained the deflection angle as 
\begin{equation}
\alpha = \pi \left\{\sqrt 
\frac{2 (r_0^2 + a^2)}{2 r_0^2 + a^2} -1 \right\},
\label{alpha-DS}
\end{equation}
where $r_0$ is the closest approach of the light.
In the weak field approximation ($a \ll b \sim r_0$), 
the deflection angle is expanded as 
\begin{equation}
\alpha = \frac{\pi}{4} \left(\frac{a}{r_0}\right)^2 
- \frac{5 \pi}{32} \left(\frac{a}{r_0}\right)^4 
+ O\left(\frac{a}{r_0}\right)^6 . 
\label{alpha-DS2}
\end{equation}
The deflection angle derived in this paper is based on 
not the closest distance but the impact parameter. 
In terms of the impact parameter, 
Eq. (\ref{alpha-DS2}) is rearranged as 
\begin{equation}
\alpha(b) = \frac{\pi}{4} \left(\frac{a}{b}\right)^2
+ \frac{3 \pi}{32} \left(\frac{a}{b}\right)^4
+ O\left(\frac{a}{b}\right)^6 . 
\label{alpha-DS3}
\end{equation}
where we used $r_0^2 = b^2 - a^2$. 

In the rigorous treatment without using the PMS approximation, 
we have obtained Eq. (\ref{alpha2}), 
the expansion of which in the weak field is given by 
Eq. (\ref{alpha3}) and explicitly written as 
\begin{equation}
\alpha(b) = \frac{\pi}{4} \left(\frac{a}{b}\right)^2
+ \frac{9 \pi}{64} \left(\frac{a}{b}\right)^4 
+ O\left(\frac{a}{b}\right)^6 . 
\label{alpha-approx}
\end{equation}
Comparing Eq. (\ref{alpha-approx}) with Eq. (\ref{alpha-DS3}) 
shows that 
the deflection angle recently expressed by Eq. (\ref{alpha-DS}) 
is valid at the leading order in the weak field approximation 
but it breaks down at the next order. 
Note that the complete elliptic integral of the first kind 
cannot be expressed by a square root like Eq. (\ref{alpha-DS}). 

Why does the previous approach fail? 
The main reason is a difference between a black hole spacetime 
and a wormhole. 
The Schwarzschild spacetime has a singularity at $r=0$, 
which also leads to a singular behavior of the light bending. 
Therefore, the PMS approximation using the delta function works 
\cite{Amore, Amore2}. 
On the other hand, $r=0$ in the Ellis geometry is a regular sphere 
which can connect with a separate spatial domain. 
The deflection angle by the Ellis wormhole is not inversely 
but logarithmically divergent there. 
Therefore, the PMS does not seem to be suitable for this case. 
Let us consider a case that the closest approach vanishes, 
for which $r_0 = 0$, namely $b=a$. 
Then, we obtain 
\begin{eqnarray}
\alpha(a) &=& 2 \int_0^1 \frac{dt}{1-t^2}
- \pi 
\nonumber\\
&\sim& \ln\infty . 
\label{alpha-a}
\end{eqnarray}
On the other hand, Eq. (\ref{alpha-DS}) leads to 
$\alpha \to \pi (\sqrt2 -1)$ as $r_0 \to 0$ ($b \to a$). 
This result misses the throat effects 
and thus it is incorrectly finite. 

Note that the throat $r=0$ is a light sphere (photon sphere). 
A light ray can stay on this sphere if it is tangential 
to the sphere, because $r=0$ satisfies Eq. (\ref{orbit}). 
The existence of the light sphere is reflected by 
the divergence in Eq. (\ref{alpha-a}).

\section{Conclusion}
%\noindent \emph{Conclusion.--- }
The light deflection by an Ellis wormhole has been reexamined. 
The bending angle as a function of the ratio between 
the impact parameter and the throat radius of the wormhole 
has been obtained in terms of a complete elliptic integral 
of the first kind. 
The deflection angle in this geometry in a different form 
\cite{Perlick, Nandi} is the same as the present one 
but it is depending on the closest approach. 
In the weak field approximation, 
it has been shown that another expression for the deflection angle 
derived (and used) in recent papers \cite{DS, BP, Abe, Toki} 
is correct at the leading order 
but it breaks down at the next order 
because there exists a throat in the Ellis geometry.

We would like to thank F. Abe and V. Perlick 
for the stimulating comments. 
This work was supported in part (H.A.) 
by a Japanese Grant-in-Aid 
for Scientific Research from the Ministry of Education, 
No. 21540252.


\begin{thebibliography}{99}
\bibitem{Nandi-2}
Eq. (54) in \cite{Nandi} uses an uncommon notation. 
Their argument of the first kind Elliptic integral 
is $k^2$ that is a square of the usual argument for $K(k)$.
\end{thebibliography}
\end{document}